# Versatile strain relief pathways in epitaxial films of (001)-oriented PbSe on III-V substrates

Brian B. Haidet[1], Jarod Meyer[2], Pooja Reddy[2], Eamonn T. Hughes[1], and Kunal Mukherjee[2*]
[1] Materials Department, University of California Santa Barbara, Santa Barbara, CA, 93106, USA
[2] Department of Materials Science and Engineering, Stanford University, Stanford, CA, 94305, USA

**ABSTRACT**
PbSe and related IV-VI rocksalt-structure semiconductors have important electronic properties that may be controlled by epitaxial strain and interfaces, thus harnessed in an emerging class of IV-VI/III-V heterostructures. The synthesis of such heterostructures and understanding mechanisms for strain-relief is central to achieving this goal. We show that a range of interfacial defects mediate lattice mismatch in (001)-oriented epitaxial thin films of PbSe with III-V templates of GaAs, InAs, and GaSb. While the primary slip system {100}<110> for dislocation glide in PbSe is well-studied for its facile glide properties, it is inactive in (001)-oriented films used in our work. Yet, we obtain nearly relaxed PbSe films in the three heteroepitaxial systems studied with interfaces ranging from incoherent without localized misfit dislocations on 8.3% mismatched GaAs, a mixture of semi-coherent and incoherent patches on 1.5% mismatched InAs, to nearly coherent on 0.8% mismatched GaSb. The semi-coherent portions of the interfaces to InAs form by 60° misfit dislocations gliding on higher order {111}<110> slip systems. On the more closely lattice-matched GaSb, arrays of 90° (edge) misfit dislocations form via a climb process. The diversity of strain-relaxation mechanisms accessible to PbSe makes it a rich system for heteroepitaxial integration with III-V substrates.

## I. INTRODUCTION

IV-VI rocksalt narrow bandgap semiconductors have long been important materials in infrared optoelectronics [1–3] and thermoelectrics, [4,5] and more recently have received attention as topological crystalline insulators [6,7] and for spin qubits [8] as a part of the broader class of quantum materials. Part of the attraction of IV-VI materials for infrared optoelectronics comes from an uncommon mix [9] between metallic, covalent, and ionic bonding, or perhaps even a distinct bonding character, [10] that brings about novel properties such as high static dielectric constants, high refractive indices, and low Auger recombination. [3,11–14] Due to the unique bonding and low growth temperatures necessary for epitaxy, semiconductors like PbSe are quite different from predominately covalently bonded zincblende III-V semiconductors like InAs despite similar bandgaps.

Device fabrication with single crystal PbSe traditionally requires epitaxy on expensive and fragile native IV-VI substrates or hygroscopic and electrically insulating $BaF_2$ substrates, both also poor conductors of heat. [15] These issues with substrate scalability and properties have renewed interest in epitaxial integration of PbSe and related IV-VI alloys with commercially available III-V, II-VI, Si, and Ge substrates. [16–21] IV-VI/III-V heterostructures not only provide a potential path to epitaxial films on large area substrates and materials already used in optoelectronics, but also provide new ways to manipulate and control electronic properties using strain and charge at the heterointerface. [18,22] Fabricating a class of devices that harness the combined properties of IV-VI and III-V materials requires an in-depth understanding of the defects that form due to the lattice constant, thermal expansion, crystal structure, and bonding mismatch between these two classes of materials.

Dislocations are the most common defect for accommodating lattice and thermal expansion mismatch strain in epitaxial systems, but dislocations in PbSe behave differently than dislocations in other rocksalt materials like NaCl due to a significantly higher lattice polarizability. [23] Most notably, the primary slip system in PbSe is {100}<110> instead of the more typical {110}<110> slip system of NaCl, with important implications for epitaxy. [24] An overwhelming majority of IV-VI heteroepitaxial growth has been geared towards (111)-oriented films prepared on (111)-oriented cubic substrates like $BaF_2$ or Si where the (100) glide planes are inclined and capable of relieving mismatch strain. Films of reasonable quality may be achieved with both lattice mismatch and thermal-expansion mismatch via (111)-oriented growth. On the other hand, the primary slip system feels no resolved shear for in-plane strain on a (001)-oriented film. Hence, the most technologically significant orientation in III-V and SiGe epitaxy is also the orientation that has no conventional way to relax thermal-expansion-mismatch or lattice-mismatch strain in IV-VI films. Thermal-expansion mismatch is particularly difficult to manage in the IV-VI/III-V (001)-oriented heteroepitaxial system (e.g. $\alpha_{PbSe}$=19 ppm/K and $\alpha_{GaAs}$=5.8 ppm/K at 300K) and the post-growth cooldown ultimately leads to cracking in thick PbSe. [25]

We still wish to leverage the high-quality surfaces and technological relevance that come with thin (001)-oriented layers on (001) III-V substrates. In this work, we show how PbSe thin films beyond the critical thickness for strain relief via dislocation formation but below the critical thickness for cracking accommodate lattice-mismatch strain during cube-on-cube epitaxy on (001)-oriented III-V substrates of GaAs, InAs, and GaSb with starting lattice constant mismatch at growth temperature of 8.3%, 1.5%, and 0.8%, respectively. We identify unusual defects that mediate the mismatch between these two crystal structures and demonstrate that PbSe is a versatile material with secondary strain relaxation mechanisms for achieving good quality thin films even when the primary slip system is not active. A more complete understanding of strain relaxation in (001)-PbSe not only facilitates integration schemes for thin films with commercially available substrates, but also potentially enables new means to tune electronic properties with atypical interfacial structures.

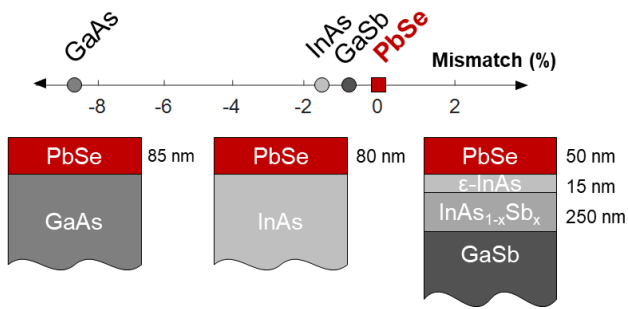

**Fig. 1.** Sketch of the three PbSe/III-V epitaxial samples in this study. Each nucleation surface is arsenic-terminated to facilitate single-crystal growth, but each surface also has a unique lattice constant. PbSe layers on GaAs, InAs, and InAsSb-on-GaSb templates range from 8.3% to 0.8% compressive mismatch.

## II. METHODS

The PbSe samples on various III-V templates in this study were synthesized using solid-source molecular beam epitaxy (MBE). Figure 1 shows the basic structure of these samples alongside the lattice-mismatch. (001)-oriented, nominally on-axis III-V substrates of GaAs, InAs, and GaSb were prepared prior to PbSe growth in a Veeco Gen III MBE system. The substrate preparation involved oxide desorption under As or Sb overpressure, followed by deposition of a homoepitaxial layer. While we have previously grown PbSe directly on GaSb(001), the film had multiple oriented nuclei and a somewhat diffuse heterointerface. [17] Therefore, in this work an additional 300 nm thick epitaxial $InAs_{0.84}Sb_{0.16}$ layer followed with a very thin layer of strained InAs was deposited on the GaSb substrate as rapidly as possible, at the ternary deposition conditions, to seal the more reactive Sb species below a less reactive surface. This allowed us to study strain relaxation close to the GaSb lattice constant while preserving an InAs-like surface chemistry that consistently yields purely (001)-oriented PbSe films. The III-V templates were finally arsenic-capped and transferred out of vacuum for PbSe growth. PbSe films of 50–80 nm thickness were deposited on GaAs, InAs, and InAs/InAsSb/GaSb templates using a Riber Compact 21 MBE system. After desorbing the arsenic cap, the III-V templates were exposed to PbSe flux at 400 °C for 20-30 seconds to prepare the surface for subsequent nucleation and growth of PbSe at 320 °C and a growth rate of 2-3 nm/minute. [17] Only a single compound effusion cell was used for PbSe, which likely results in Pb-rich n-type thin film samples. In all cases, RHEED appears streaky across the nucleation step, but we have previously noted during growth on GaAs and InAs substrates that the growth mode is still of the Volmer-Weber island type, just with very flat (001)-oriented islands. [17]

The in-plane and out-of-plane lattice parameters and film morphology are determined using coupled 2θ-ω scans, reciprocal space maps (RSMs), and transverse scans collected using triple-axis x-ray diffraction on a Panalytical X'Pert instrument. The transverse scan is like a rocking curve measurement but uses the monochromator on the detector side, as opposed to a double-axis scan with a wide-open detector for a classical rocking curve. [26] In the case of large and moderate mismatch with GaAs and InAs, we focus primarily on the film morphology determined by the transverse scans, the atomic arrangement at the interface, and dislocation network as much of the starting mismatch strain is relaxed even for ultrathin films. On the other hand, we use x-ray reciprocal space maps (RSMs) to more accurately study strain relaxation in PbSe on the low-mismatch InAsSb/GaSb template. We note that although PbSe has a larger bulk lattice parameter than all the substrates studied here, we find the film tensile strained at room temperature due to a large thermal expansion mismatch between the PbSe and the III-V substrate. We can ignore this thermal mismatch in the analysis of relaxed strain as we assume thermal expansion strain is unrelaxed during cool down in our thin films.

The atomic arrangement at interface and dislocations are analyzed further in cross-section using a TFS Talos scanning transmission electron microscope (STEM) operating at 200 kV. A high-angle annular dark field (HAADF) detector was used to collect image sequences at the PbSe/III-V interface. These sequences were then drift-corrected and stacked to resolve atomic columns. Individual atomic column positions were measured by taking 1D line traces parallel to the growth surface, convolving these traces with a Gaussian curve representative of a single atomic column, and locating peaks in the resulting smoothed signal. Dislocations in the sample on InAsSb/GaSb are additionally characterized in plan-view using electron channeling contrast imaging (ECCI) in an Apreo-S scanning electron microscope (SEM) at 30 kV.

## III. RESULTS
### III.A. Epitaxy on highly mismatched GaAs substrates
We find that PbSe grows epitaxially on GaAs (001) with a conventional cube-on-cube orientation despite a severe 8.3% compressive lattice mismatch at a growth temperature of 320 °C. This agrees with recent work showing (001)-oriented PbSe films on GaAs with a different nucleation method. [18] We have shown previously that PbSe nucleates as islands on the substrate surface that eventually coalesce into a contiguous film. [17] Prior to coalescence, these PbSe islands on GaAs are faceted with non-polar low energy {100} surfaces and edges, and after coalescence very flat (001) surfaces are typical. Figure 2a shows triple-axis 2θ-ω coupled scans of the (002) reflection from an 85 nm film of PbSe on GaAs. Pendellösung fringes indicate a sharp interface. The out-of-plane lattice parameter derived from this scan suggests the film is only about 0.1–0.2% tensile strained in-plane, relieving nearly all the compressive lattice-mismatch strain during growth. Note the measured strain at room temperature is complicated by a buildup of tensile strain (< 0.3%) induced by thermal expansion mismatch between PbSe and GaAs during cooldown, but regardless, any strain relief mechanism that occurs during growth accommodates about 8% mismatch.

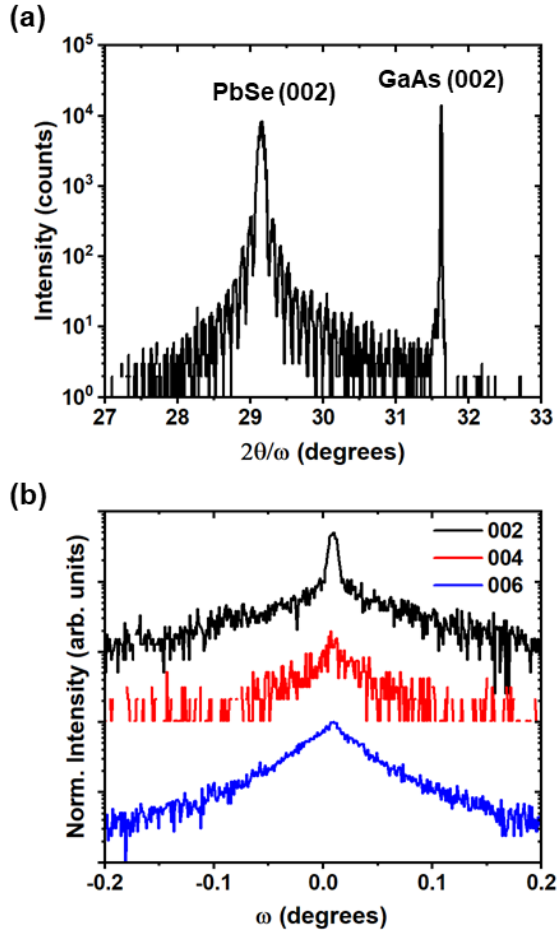

**Fig. 2.** (a) Symmetric scans of the (002) reflection of PbSe thin films of varying thickness on GaAs substrates, collected in a triple-axis geometry. The film thickness is obtained via fitting to the fringing (b) A transverse (triple-axis) rocking curve of the (002), (004), and (006) reflections showing a classic two-component peak: an instrument-resolution limited coherent Bragg reflection above a defect-broadening peak. The intensity of the coherent reflection reduces drastically upon increasing the magnitude of the scattering vector.

Figure 2b shows a transverse scan for this film for the (002), (004), and (006) reflections of PbSe. Interestingly, the (002) transverse scan resolves both a narrow and a wide component, the former barely visible in the (004) reflection and absent in the (006) reflection. The narrow peak is also absent for all reflections in the conventional open detector rocking curve scan. Miceli et al. first discussed such a two-component reflection in ErAs/GaAs films (coincidentally rocksalt on zincblende). [27] They propose that the narrow component at the center is an instrument broadening limited peak that corresponds to a specular reflection or coherent scattering from long-range ordering of atoms in the film, while the second wider peak is the typical diffuse scattering arising from short-range correlations in a mosaic-structured film. [28] Later, other groups placed this interpretation on firmer theoretical grounds on the basis of interfacial misfit dislocation networks. [29,30] Two-component transverse scan signatures have since been observed in a range of epitaxial films, spanning metals to ceramics. [26,31]

The intensity of the coherent scatter component is predicted to reduce with increasing order (hkl) of the reflection, also seen in our experiments (and more clearly in section III.B), but this expectation is for very thin films whose thickness is on the order of the average misfit dislocation spacing. [30] Remembering that the favored Burgers vector of dislocations in PbSe is $\frac{a}{2}\langle 110 \rangle$ (similar to III-V zincblende materials), a semi-coherent interface with a square network of edge misfit dislocations to relieve 8% strain would correspond to a dislocation spacing of approximately 5 nm, much smaller than the film thickness. More recent modeling, however, predicts the coherent scatter component can remain strong in films much thicker than the average misfit dislocation spacing, but only if the misfit dislocations themselves are well ordered. [32] Therefore, the fact that we see the coherent peak even in an 85 nm thick film suggests a high degree of order or periodicity in the defect structure at the interface.

Cross-sectional STEM imaging of this film of PbSe on GaAs (Fig. 3a) shows a columnar morphology evidenced by numerous vertical features corresponding to either threading dislocations or low-angle grain boundaries—defects expected from island nucleation and coalescence. Fig. 3b shows a magnified view of the interface between PbSe and GaAs, revealing a somewhat ordered defect structure corresponding to the predicted 5 nm period. Remarkably, the higher magnification Fig. 3c reveals this defect structure is unlike a conventional array of misfit dislocations. We find the periodic structure corresponds to a rippling of the initial PbSe layer and/or the final layer of the arsenic-terminated zincblende surface. A Burgers circuit reveals a net displacement across the interface around each of these ripple features, fulfilling the strain relaxation function of a dislocation, yet the atoms are uniformly distributed without any local strain of a typical dislocation core. The displacement of 13 atomic columns of arsenic corresponds with the displacement of 12 atomic columns of Pb (7.6% strain relief), mirroring a conclusion from Liu et al. on PbSe/GaAs, although they do not specifically note ripples in their images. [18] It is possible that differences in PbSe nucleation strategies between their work and ours (Se-exposure vs. PbSe-exposure) leads to this. Fig. 3d plots the measured disregistry between the sets of atom columns at two ripples in the image (labeled TEM-1 and TEM-2), and compares to analytical predictions of disregistry derived from a modified Peierls-Nabarro model developed by Yao et al. for interfacial misfit dislocations between two materials with lattice parameters $a_1$ and $a_2$. [33] Three cases of disregistry for different values of a dimensionless parameter $\beta = \frac{2\pi\mu f}{\tau}$ are calculated and shown. Here, $\mu$ is a modified shear modulus ($\mu \sim \mu_{PbSe}$ in our case), $f$ is the misfit defined as $\frac{2(a_1-a_2)}{(a_1+a_2)}$, and $\tau$ is the bond strength or shear modulus parallel to the interface. We find values of $\beta > 2$ fit the measured disregistry well, and this corresponds to an interfacial bond strength $\tau < 7\ GPa$. We require more precise STEM measurements to say how much the disregistry deviates from the limiting case of a perfectly straight line ($\tau = 0\ GPa$) of an unbonded film.

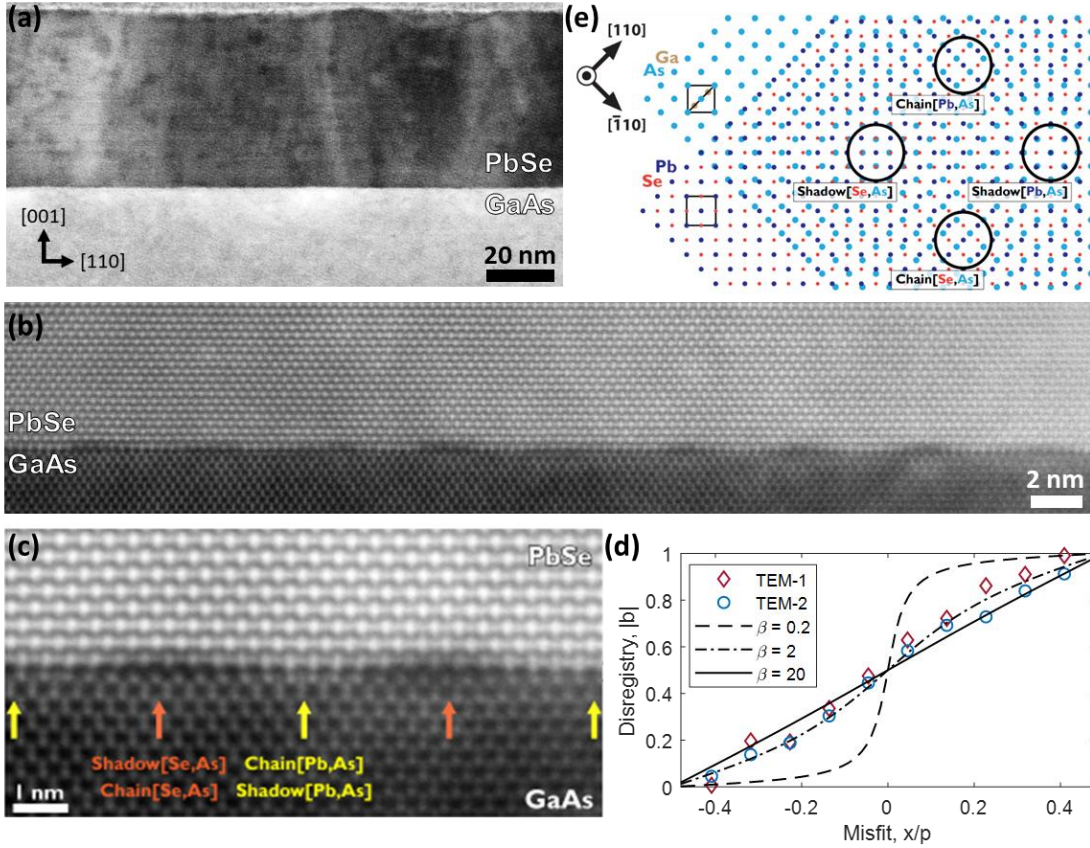

**Fig. 3.** (a) Cross-sectional STEM image of the 60 nm PbSe-on-GaAs film (b) HAADF-STEM image of the PbSe-GaAs interface along the [110] cross-section showing a periodic set of undulations, seen in higher magnification in (c). (d) The experimentally determined disregistry of atoms (first PbSe layer vs. last GaAs layer) relieving the nearly 8% mismatch shows a smooth profile. (e) Sketch, to scale, of the unrelaxed positions of atoms in a Pb-Se (001) monolayer and As-atoms of a GaAs (001) monolayer. The shifting vertical alignment between these two lattices results in so-called Chain and Shadow configurations.

It suffices to say that the PbSe/GaAs interface appears incoherent (also called incommensurate) without condensed misfit dislocation cores and implies weak bonding between the PbSe film and GaAs substrate underscored by the low value of $\tau$. We propose that the unusual structure of the interface, where the lateral periodicity of atoms is maintained over large distances instead of being disrupted by a randomly spaced network of misfit dislocations, results in the intense "coherent component" observed in the x-ray transverse scans even for a thick film. We currently have no view of the defect structure at island coalescence boundaries in PbSe/GaAs but suspect there may be conventional misfit dislocations at the interface near these features.

In early work on rocksalt/zincblende epitaxy, Tarnow coined the terms 'shadow' and 'chain' to specify, for ErAs/AlAs interfaces, whether the atom in the first rocksalt layer was directly above, or shifted with respect to the terminating zincblende atom, respectively. [34] In that work, it was assumed correctly that ErAs had a consistent alignment to the AlAs underlayer when far from condensed dislocation cores, typical behavior at most mismatched interfaces. In our work, we observe highly unusual ordering – there is not a single chain or shadow configuration for the PbSe/GaAs interface – the stacking sequence varies continuously in space due to the smooth disregistry. Fig. 3e is a plan-view sketch of the moiré pattern formed by superimposing the final As-terminated (001) zincblende surface and the first rocksalt PbSe (001) surface. To approximate the incommensurate interface, both lattices are drawn unrelaxed, with uniform in-plane lattice constants. We see that the interface smoothly transitions between four different stackings: chain and shadow configurations where both pairs of atoms interact across the interface. Based on atomic number sensitive HAADF contrast, we speculate that the interfacial rippling seen in Fig. 3c arises due to electrostatic repulsion between the more electronegative Se and As atoms in the chain or shadow configuration, and/or electrostatic attraction between Pb and As atoms in the chain or shadow configuration. As the deregistered interface brings different pairs of atoms into proximity laterally, they respond with out-of-plane displacement. The properties of this complex interface remain to be explored, but we note that preliminary work on PbSe/GaAs samples exhibit surprisingly long minority carrier lifetimes even at room temperature. [35] It is possible that the absence of localized misfit dislocations plays a role in enabling this.

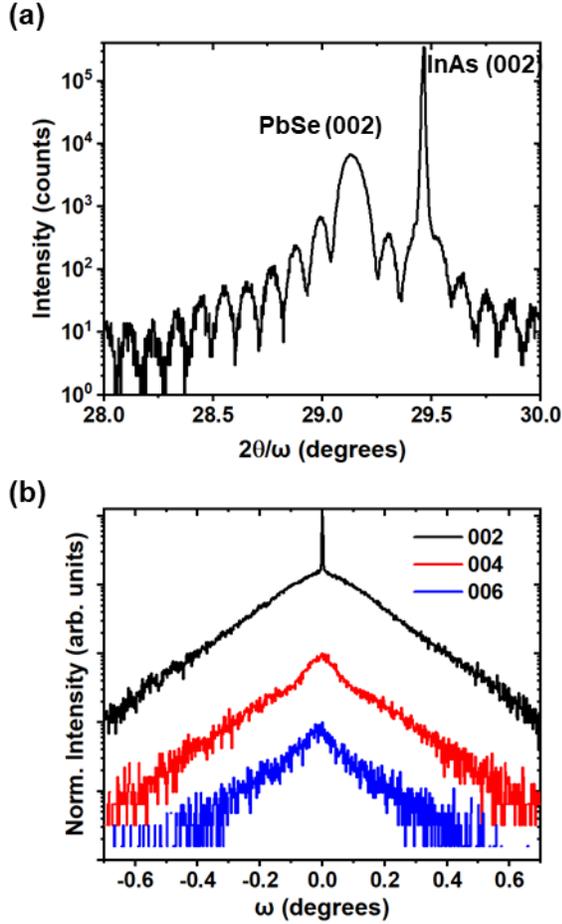

**Fig. 4.** (a) Symmetric scans of the (002) reflection of the 80 nm PbSe thin film on InAs substrate, collected in a triple-axis geometry. (b) A transverse (triple-axis) rocking curve of the (002) reflection showing a two-component peak for the (002) reflection: an instrument-resolution limited coherent Bragg reflection above a defect-broadening peak, the intensity of the coherent reflection disappears upon increasing the magnitude of the scattering vector.

**III.B. Epitaxy on moderately mismatched InAs substrates**

PbSe is 1.5% compressively mismatched to InAs at a growth temperature of 320 °C. Despite the lower mismatch, layer-by-layer nucleation does not occur, and Haidet et al. show PbSe nucleates on InAs(001) as islands with (001) surfaces before coalescing into a continuous film. [17] A key difference in island morphology relative to growth on GaAs is that the islands are elongated and have {110} edges rather the natural low energy {100} facets of PbSe. [17] This suggests more interaction with the III-V substrate, possibly a result of increased interfacial bond strength or surface-reconstruction-mediated adatom mobility. Fig. 4a shows triple-axis 2θ-ω coupled scan of the (002) reflection of the 80 nm film showing good interface sharpness with several Pendellösung fringes. The out-of-plane lattice spacing of the PbSe film reveals a strain of less than ±0.02%, that the film is almost fully relieved of its 1.5% in-plane lattice-mismatch strain. A transverse scan of the 80 nm PbSe film on InAs reveals a clear two-component peak for the (002) reflection that is absent for the higher order reflections, like that seen previously on GaAs.

The cross-sectional STEM micrograph of the PbSe/InAs interface in Fig. 5a resolves distinct misfit dislocations spaced about 40 nm, made visible using diffraction contrast and by tilting the foil by a few degrees about the horizontal. Figs. 5b-c show a high-resolution view of two such misfit dislocations revealing conventional compact cores with a Burgers vector (Burgers circuit shown using a RH-FS convention) of type $\vec{b} = \frac{a}{2}\langle 110 \rangle$ that has an in-plane component of $\frac{a}{4}\langle 110 \rangle$ relieving compressive strain. The misfit dislocations are 60° dislocations commonly seen in low mismatch strain relief. However, we see this network results only in partial relief of 0.7% compressive strain out of a total 1.5% with this magnitude of Burgers vector and dislocation spacing. Upon further inspection of the interface, we observe another type of strain relieving defect that resembles the interface structure observed at the PbSe/GaAs interface. We show in Fig. 5d a section of the interface that contains the Burgers vector content of an edge dislocation with $\vec{b} = \frac{a}{2}\langle 110 \rangle$, relieving compressive strain of 2% (displacing the film by a net of 1 atomic column over 50 atomic columns with respect to the substrate). Remarkably, the disregistry between PbSe and InAs is distributed uniformly to within the limits of our measurement across almost 50 atomic columns—very unusual for low mismatch films. Figure 5e plots this measured disregistry alongside calculated misfit dislocation disregistry for different values of β in the modified Peierls-Nabarro model mentioned previously. We find that the PbSe/InAs interface strength of these incoherent sections correspond to $\tau < 1\ GPa$. We conclude strain relaxation in PbSe/InAs is accomplished together by localized dislocations and incoherent interface sections. We think the periodically spaced dislocations in the semi-coherent sections and the linear disregistry in the incoherent sections yield the clear coherent component peak in the XRD transverse scan of Fig. 4b.

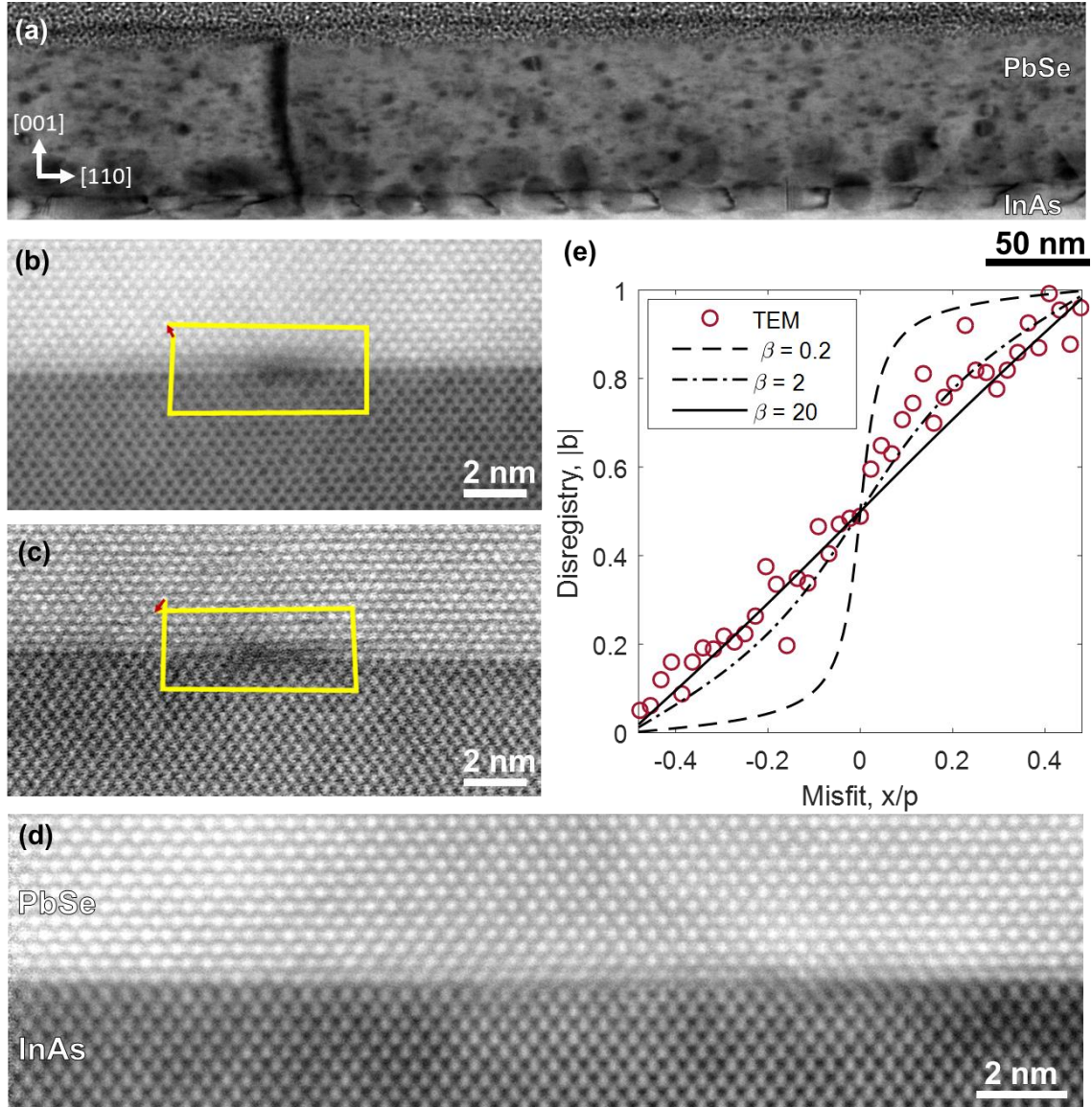

**Fig. 5.** (a) Cross-sectional STEM image of the 80 nm PbSe-on-InAs film. The foil is tilted slightly along the horizontal to better see the array of misfit dislocations at the interface. (b,c) Higher magnification HAADF-STEM images of the [110] cross-section showing the core of these misfit dislocations. A Burgers circuit reveals these dislocations relieve compressive mismatch with an in-plane component of a/4[110]. (d) Wide view of adjacent sections of this PbSe-InAs interface collected using HAADF-STEM, with net compressive-mismatch relieving Burgers vector content of a/2[110] without localized dislocation cores. An irregular partial layer of atom columns is seen as the interface transitions to the shadow/chain [Se,As] configuration. (e) The experimentally determined disregistry of atoms (first PbSe layer vs. last GaAs layer, excluding the anomalous partial layer of atoms) showing a nearly linear behavior indicative of an incoherent interface.

We had previously characterized the PbSe/InAs interface as chain [Pb,As]. [17] This is indeed true near localized misfit dislocations. Yet, in sections where we find an incoherent interface (see Fig. 5d), we note an additional partial row of atoms in between sections of registry, modifying this stacking. The partial row of atoms accommodating an otherwise linear disregistry exists in regions where [Se-As] alignment occurs. In PbSe/GaAs, we observe rippling at these sites. The much wider section of this overlap in PbSe/InAs may place some cations (Pb or In) in such a layer to mediate electrostatic repulsion between As and

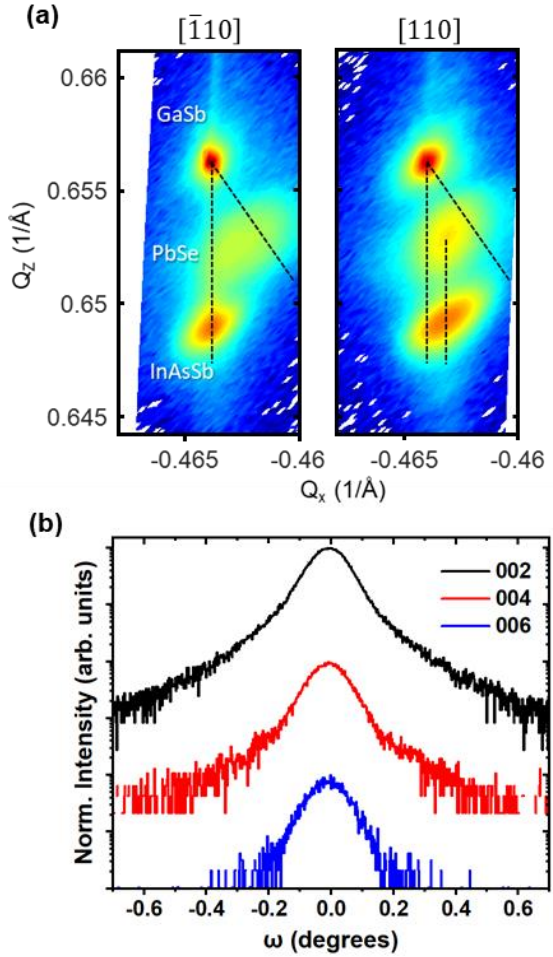

**Fig. 6.** (a) Reciprocal space maps of the {224} reflection of a PbSe thin film on InAsSb-on-GaSb substrate collected along the [$\bar{1}10$] and [110] azimuths, respectively. The axes are shown relative to the GaSb substrate. The InAsSb peak is coherently strained to the GaSb substrate along [$\bar{1}10$] and partially relaxed via misfit dislocations along the [110] direction. In turn, the PbSe is coherent to InAsSb along the [110] and partially relaxed along the [$\bar{1}10$] with respect to InAsSb. The tilted dashed line shows the fully relaxed condition. (b) A transverse scan of a set of symmetric reflections from PbSe on InAsSb/GaSb, none of which show the coherent Bragg reflection.

Se. In summary, the interface between PbSe and InAs is no less complex than GaAs and is an important platform to understand and control. Because these two materials have similar bandgaps and high spin-orbit coupling, the interfacial structure and charge and the resulting changes to band-alignment may be important for interfacial tunneling phenomena among others.

### III.C. Epitaxy on low-mismatch InAsSb/GaSb templates

GaSb is a wider bandgap substrate than InAs also with the potential for high-quality IV-VI epitaxy thanks to only a 0.8% smaller lattice constant than PbSe at the growth temperature. [36] Yet, initial attempts at direct growth on GaSb(001) desorbed under Sb flux results in minority misoriented (221) or (110) nuclei of PbSe in addition to the majority (001)-oriented nuclei potentially due to Sb-Se interactions. [17] This precludes a direct comparison to GaAs and InAs. To achieve single orientation nucleation at this lattice constant, we grew a 250 nm InAsSb buffer layer capped with 15 nm of InAs on a GaSb(001) substrate to preserve an arsenic-rich surface chemistry without significantly altering the buffer lattice constant. RSM analysis in Fig. 6a shows the InAsSb buffer is only partially and asymmetrically relaxed, an unintended consequence of our growth parameters but leading to interesting results. The buffer is nearly coherent to GaSb along the [$\bar{1}10$] direction (seen in the identical values of the in-plane spacing $Q_x$), presenting a lattice-mismatch of 0.8% to PbSe at growth temperature. We see that a 50 nm thick PbSe film grows partially relaxed along this azimuth with a 0.13% larger in-plane lattice parameter to InAsSb. Along the [110] direction, however, the InAsSb template itself is partially relaxed with respect to the GaSb substrate and presents a lower in-plane lattice mismatch to PbSe of 0.67%. As seen by identical $Q_x$ values, we find PbSe coherently strained to the InAsSb template along this closer matched [110] direction. A transverse scan of the PbSe film shown in Fig. 6b no longer shows the coherent scatter component of the peak even for the lowest (002) reflection. This suggests a fundamentally different nature of order in the PbSe lattice.

We use a combination of ECCI and STEM to study defects facilitating PbSe relaxation for this low-mismatch case. Figure 7a shows a plan view ECCI image collected in a multi-beam condition ($g = [\bar{2}20]$ and $g = [040]$) showing an unusual one-dimensional array of misfit dislocations with an average spacing of 90 nm arranged vertically in the figure. Typically, we are unable to use ECCI to see (localized) interfacial misfit dislocations in the samples described previously due to the limited depth sensitivity of the technique as well as the close spacing of the misfit dislocations. A **g.b** analysis shown in Fig. 7b suggests the vertical misfit dislocation segments are edge type, disappearing when $g = [\bar{2}\bar{2}0]$ or $g = [220]$, and appearing (with opposite contrast) when $g = [\bar{2}20]$ or $g = [2\bar{2}0]$. We can see in Fig. 7a threading dislocations bound the edge misfit dislocations. These threading dislocations ends organize in rows or bunches, perhaps linked to areas of roughness. Cross-sectional STEM of the PbSe layer shown in Fig. 7c confirms the existence of a periodic array of misfit dislocations only along one direction, situated, remarkably, midway into the PbSe film thickness rather than at the PbSe/III-V interface. Fig. 7c shows HR-STEM of one of these misfit dislocations confirming an edge character with an overall Burgers vector of $\frac{a}{2}\langle 110 \rangle$ that relieves compressive strain. We discuss in section IV.C the origins of these misfit dislocations via a climb mechanism. If we take a closer look at the position of the atoms around the misfit dislocation in Fig. 7d, we find that the edge dislocation is split into two partial dislocations separated by a 2 nm wide stacking fault (Fig. 7e., red double arrow). Yet, the disregistry of atoms above and below the partial dislocation cores remains localized to a few atoms spacing. This finding is unsurprising in a bulk material like PbSe where interatomic bonds are strong.

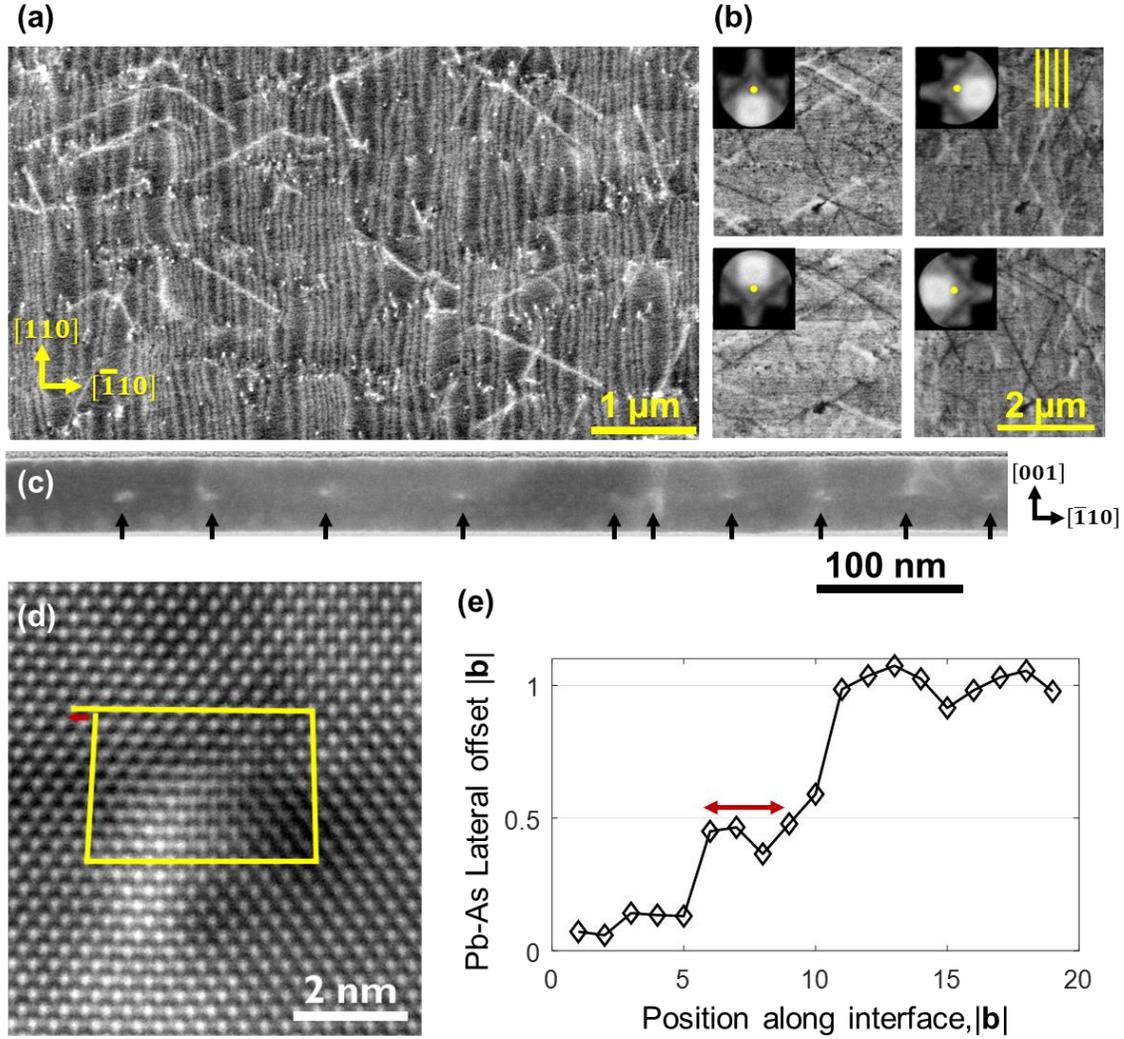

**Fig. 7.** (a) A unidirectional array of misfit dislocations in PbSe-on-InAsSb/GaSb seen via ECCI. The misfit dislocations terminate on the film surface via threading dislocations, visible as white dots. A smaller set of misfit dislocations with unusual line directions are also visible (b) A set of images of this array of misfit dislocations collected using $g = (220), (\bar{2}\bar{2}0), (\bar{2}20),$ and $(2\bar{2}0)$, with the misfit dislocations disappearing with $g \cdot b = 0$, suggesting pure edge dislocations with $\vec{b} = \frac{a}{2}[\bar{1}10]$. (c) Cross-sectional image of the PbSe layer cut along the [110] showing the array of misfit dislocations existing mid-way through the film. (d) Higher magnification image of the core structure of the misfit dislocation, confirming the edge character of the defect. (e) The disregistry, measured as the Pb-As lateral offset, shows the edge dislocation is split into two partial dislocations separated by a stacking fault (red arrow). Both cores are condensed and accommodate the disregistry within a few b-vector spacing.

Correlating the defect structure to the strain relaxation in PbSe on InAsSb/GaSb, we find the anisotropy of the array of edge misfit dislocations agrees with the relaxation azimuth seen in the RSM. The in-plane strain relaxed by the dislocation array spaced 90 nm corresponds to approximately 0.24% in the upper half of the PbSe film with respect to the lower half. Thus, the film-average relaxation of 0.13% measured by RSM suggests that the lower half of the PbSe film is fully coherent (0% strain relaxed). The ECCI image of Fig. 7 also shows a sparse set of misfit dislocations with crystallographic line directions, although the glide plane is not clear to us without further analysis. The low density of these defects is in line with a nearly coherent PbSe/III-V interface along the [110] direction. In summary, the occurrence of a high density of threading dislocations in the heteroepitaxy of PbSe on III-V substrates is attributed to a combination of island growth and coalescence due to lack of wetting of the substrate in addition to any lattice-mismatch. [17] By using a very closely lattice-matched InAsSb buffer, we can show that at least in the [110] direction, islands are able to coalesce without defects. This is promising towards developing defect-free PbSe/III-V integration using a symmetrically relaxed, lattice-matched buffer.

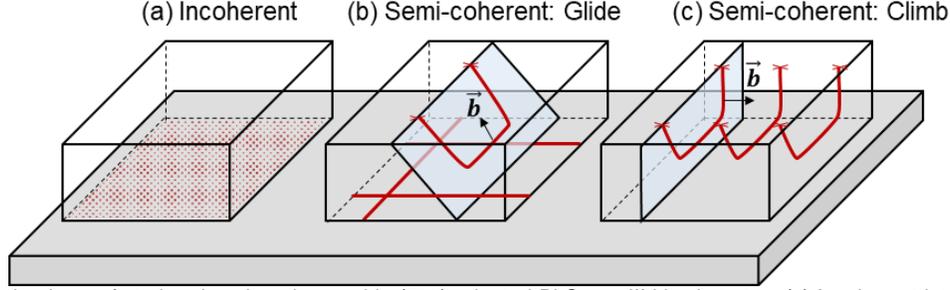

**Fig. 8.** Three mechanisms of strain relaxation observed in (001)-oriented PbSe on III-V substrates. (a) Incoherent interface on GaAs and InAs, (b) Semi-coherent system via {111}<110> dislocation glide on InAs, and (c) Semi-coherent system via unidirectional edge dislocation array formed by climb on InAsSb/GaSb.

## IV. DISCUSSION
### IV.A. Incoherent PbSe/III-V interfaces and transition to semi-coherence

We show that (001)-oriented PbSe films harness a range of defects to mediate lattice-mismatch on three arsenic-terminated III-V substrates of varying mismatch, despite no access to its primary slip system. Depending on the choice of substrate, PbSe can form interfaces spanning incoherent, semi-coherent, and (nearly) coherent structures, sketched in Fig. 8.

This gradual transformation in interface character is a trade-off between interface energy and long-range strain in the film. A low energy coherent interface (hence high film-substrate adhesion strength) forms in films with low mismatch, but the substrate strains the film fully. Such a state is thermodynamically preferred up to a limit. [37] Beyond this, such as for intermediate mismatch and a reasonable thickness of film, the interface develops an array of misfit dislocations with condensed cores that partially relieves strain in the film, while sections of the interface between dislocations remain in registry and atom columns across the dislocation cores are deregistered—this is the semi-coherent state. [38] GaSb-on-GaAs with a 7% mismatch [39] or even InAs on GaP with a 11% mismatch [40] are examples of such an interface where condensed dislocation cores are periodically spaced and the regions in between the dislocations are in perfect registry as seen in TEM. If the mismatch is increased further, say to more than 10–15%, the condensed dislocations are spaced closer and the disregistry across them start to overlap and hence it is not reasonable to describe the interface as an array of misfit dislocations anymore and the interfaces is often referred to as being incoherent. [38] Incoherent interfaces with orientation relationships, often seen in metal-ceramic interfaces, are energetically expensive and are associated with weak bonding between the film and substrate. [41–44]

By this description of incoherent interfaces, it is very unusual to see them in the lower mismatched films (<10%) on GaAs and InAs. Thus, we can now appreciate that the absolute value of mismatch alone does not dictate whether an interface will be incoherent or semi-coherent. The nature of bonding of the interface is also important. [45,46] Romanov et al. modeled a transition of an interface from semi-coherent to incoherent as a delocalization or spreading of misfit dislocation cores and proposed a metric based on energy minimization that only depends on some material parameters for when this transition may occur. [47] They suggested that the value of a parameter $C = \frac{4\pi\Delta\gamma}{\mu|f|b}$ might indicate appreciable amounts of incoherent patches of an interface when $C < 10$. Here $\Delta\gamma$ is a phenomenological interface energy term that represents the difference in interface energy between the coherent and incoherent state due to changes to short-range chemical forces (and does not contain elastic energy terms). This value is low for a weakly bonded interface, like in a vdW heterostructure, as there is little difference between the two configurations. $f$ is the misfit defined in III.A, and $\mu$ and $b$ are the shear modulus and Burgers vector, respectively. Clearly, an interface is more likely to become incoherent as the misfit $f$ increases or $\Delta\gamma$ decreases (signifying weak bonding). We find $\Delta\gamma_{PbSe-InAs} \sim 0.4\,J/m^2$ by setting $C = 10$. At the same time, we know that PbSe/PbTe and GaSb/GaAs interfaces remain semi-coherent, with compact dislocation cores, even with $f = 0.05$ and $f = 0.08$, respectively. This sets $\Delta\gamma_{PbSe-PbTe} > 1.5\,J/m^2$ and $\Delta\gamma_{GaSb-GaAs} > 2.4\,J/m^2$. Recently, it was reported that epitaxial islands of PbTe on (111)A InP can reorient themselves during growth. [20] This result is reminiscent of observations in the metal/ceramic case with incoherent interfaces. As an example, Ni films on MgO have just enough interface bonding to set initial orientation relationships, but they are nonetheless weakly bonding as seen by a complete reorientation of the Ni film upon annealing. [48] In summary, we are finding that interfaces between PbSe and GaAs or InAs (001) are starting to resemble those in metal-on-ceramic films or vdW heterostructures. This might be the result of reduced covalent (more nondirectional) character of bonding in PbSe and related IV-VI materials.

**IV.B. Dislocation mediated strain relaxation in semi-coherent films**

We move to the conventional periodic array of misfit dislocations with compact cores in semi-coherent sections of interface in PbSe/InAs. These misfit dislocations form via operation of the higher order {111}<110> slip system in PbSe, judging by the [110] line directions of the dislocations and their 60° mixed character. Plastic deformation by dislocation glide is possible even in (001)-oriented PbSe films. [49,50] It was shown in bulk IV-VI materials that the {110}<110> and {111}<110> slip systems are active at moderate temperatures; the trend in temperatures when these secondary slip systems become active decreases with decreasing ionicity as PbS (773 K), PbSe (636 K), and PbTe (417 K). [51] We do not know why the {110}<110> slip system is not active in our experiments as it has a higher Schmid factor of 0.5 compared to 0.41 for {111}<110>. Nevertheless, the equilibrium critical thickness for strain relief by 60° misfit dislocations is approximately 80 nm for 1.5% biaxial strain, [52] which is just about the thickness of our relaxed film. This agreement suggesting relaxation just above the critical thickness points towards low kinetic barriers to dislocation glide also on the {111}<110> slip system.

The unusual one-dimensional array of compressive strain relieving misfit dislocations in the initially coherent case of 50 nm PbSe on InAsSb/GaSb template (Fig. 8c) do not form by glide. These edge dislocations have a glide plane of (001), parallel to the PbSe/III-V interface. Rather, by noting that many of the misfit dislocations connect clearly to two threading dislocation segments (Fig. 7a), we propose that they nucleate as loops on the surface and expand by a climb mechanism. That is, PbSe might nucleate on the closely lattice-matched III-V template either in layer-by-layer mode or as coalescing wide islands, following which the continuing buildup of strain energy results in dislocation loop nucleation from the surface at some critical thickness that is below 50 nm. For context, the equilibrium critical thickness at which a 0.8% biaxially strained film introduces edge-character misfit dislocations is approximately 70 nm, [52] close to our result. The critical thickness for strain relaxation increases to 100 nm for 0.67% strain and therefore the trend is in line with only partial relaxation of PbSe in one direction. If the dislocations form conventionally prior to island coalescence by injection at the island edges or by reaction of other glissile dislocations, there is no reason for them to lie mid-way through the film. The organization of the threading dislocation ends of the misfit dislocations hint to their origins in heterogeneous surface loop nucleation. Indeed, using analytical expressions for half loop formation given by Hull and Bean, an insurmountable activation barrier of nearly 400 eV (and 9 nm loop radius) is found for homogenous surface nucleation with only 0.8% biaxial strain. [53] We suspect the threading dislocations lie along step edges on the film surface where the stress may be concentrated thereby facilitating heterogeneous nucleation.

Moving from the energetics of strain relaxation to its kinetics, dislocation climb is typically much more sluggish than dislocation glide. Nevertheless, there are prior reports of dislocation climb in thin films when dislocation glide is either inactive or inefficient. The slow diffusion of point defects sets the rate of dislocation climb and may result in the incomplete strain relief (misfit dislocations midway in the film) we see. Trampert et al. see strain relief by dislocation loops that climb down from the surface when they specifically induce layer-by-layer or 2D growth of 5 monolayers of 7%-mismatched InAs on GaAs (as opposed to the conventional 3D or island growth that facilitates a glide-based strain relaxation). [54] Even in IV-VI materials, Samaras et al. invoke a climb mechanism for the formation of a square network of <110> oriented edge dislocations in the layer-by-layer growth of 20 nm films of (001)-oriented $PbSe_xTe_{1-x}$ on mismatched PbTe templates with the selenium site fraction x varying from 0.2 to 1 and the corresponding mismatch increasing from 1% to 5.2%. [55] Springholz et al. track directly via scanning tunneling microscopy the formation of this network in a few monolayers of (001)-oriented films of PbSe-on-PbTe, likewise invoking a climb mechanism. [56] The sample in our study is thicker than these previous cases, but then the strain is also much lower. We might have inadvertently accessed a climb mechanism even for low mismatch by keeping the growth front planar, thereby avoiding the formation of conventional dislocations. Perhaps the reasonably high relative temperature of growth (~40% of melting temperature of PbSe) or the large degree of nonstoichiometry [57] in the Pb-rich film permits dislocation climb to be a viable mechanism for strain relaxation. Easy experimental access to dislocation climb presents new opportunities to understand the interplay between electrically active point defects and mechanical properties.

**V. CONCLUSIONS**

The unique bonding in PbSe not only confers useful electronic properties but also uncommon structural and mechanical properties for integration with III-V substrates. We have shown that (001)-oriented PbSe epitaxial films form an incoherent interface with 8.3% lattice-mismatch (001)-GaAs with a 13/12 coincident lattice, absent of condensed dislocation cores. For 1.5% lattice-mismatch strain on InAs, PbSe harnesses a higher-order slip system {111}<110> yielding a semi-coherent interface with a periodic array of misfit dislocations. Yet unusually for this low of a mismatch, we find other sections of this interface retain an incoherent character much like that on GaAs. For the lowest level of mismatch below 1% on InAsSb/GaSb, we expect PbSe to nucleate coherently or nearly coherently, and we see a climb-driven process generate an array of misfit dislocations when the film exceeds the critical thickness for dislocation-mediated strain relief. The flexibility of bonding in PbSe underpins this diversity of strain relaxation mechanisms and makes it a versatile material for heteroepitaxial integration and strain tuning.

**Data availability**

The data that support the findings of this study are available from the corresponding author upon reasonable request.

**Acknowledgements**

We gratefully acknowledge support via the NSF CAREER award under grant No. DMR-2036520 and the UC Santa Barbara NSF Quantum Foundry funded via the Q-AMASE-i program under award DMR-1906325. X-ray diffraction was performed at the Stanford Nano Shared Facilities (SNSF), supported by the NSF under award ECCS-2026822. We acknowledge the use of shared facilities of the NSF MRSEC at UC Santa Barbara for STEM imaging under award DMR-1720256. B.B.H. acknowledges support from the NSF GRFP under grant DGE-1650114.